\documentclass[twocolumn, 
prb,preprintnumbers,amsmath,amssymb,floatfix]{revtex4}

\usepackage{graphicx}
\usepackage{subfigure}
\usepackage{dcolumn}

\begin{document}

\title{Strong Electron-Phonon Coupling in Yttrium under Pressure}
\author{Z. P. Yin, S. Y. Savrasov, and W. E. Pickett}
\affiliation{Department of Physics, University of California, Davis, California, 95616}
\date{\today}
%%%%%%%
\begin{abstract}
Linear response methods are applied to identify the increase in electron-phonon
coupling in elemental yttrium that is responsible for its high 
superconducting critical temperature T$_c$,
which reaches nearly 20 K at 115 GPa.  
While the evolution of the band structure and
density of states is smooth and seemingly  modest, there is strong increase in
the $4d$ content of the occupied conduction states under pressure.  We find that the transverse
mode near the L point of the fcc Brillouin zone, already soft at ambient
pressure, becomes unstable (in
harmonic approximation) at a
relative volume $V/V_o=0.60$ ($P \approx$ 42 GPa).  The coupling to 
transverse branches is 
relatively strong at all high symmetry zone boundary points X, K, and L.
Coupling to the longitudinal branches is not as strong, but extends 
over more regions of the Brillouin zone and involves higher frequencies.
Evaluation of the electron-phonon spectral function $\alpha^2F(\omega)$
shows a very strong increase with pressure of coupling in the 2-7 meV
range, with a steady increase also in the 7-20 meV range.  These results
demonstrates strong electron-phonon coupling in this system that can
account for the observed range of T$_c$.
\end{abstract}
\maketitle

\section{Introduction}
The remarkable discovery\cite{MgB2} in 2001 of MgB$_2$ with  
superconducting critical temperature T$_c$=40K,
and the fact that the simple free-electron-like metal 
lithium\cite{lithium1, lithium2, lithium3} 
also has T$_c$ in the 14-20K under 30-50 GPa pressure, has greatly 
increased efforts in seeking higher T$_c$ in elements and simple compounds. 
Currently 29 elements are known to be superconducting at ordinary pressure 
and 23 other elements superconduct only under pressure\cite{ashcroft, hemley}.
Among elements there is a clear trend for those with small atomic number 
Z to have higher values of T$_c$, although much variation exists.
For example, under pressure\cite{jss1,table} Li, B, P, S, Ca, and V all have 
T$_c$ in the range 11-20 K.
Hydrogen\cite{hydrogen1, hydrogen2,hydrogen3},
the lightest element, is predicted to superconduct at much higher
temperature at pressures where it becomes metallic. 

While light elements tend to have higher T$_c$ among elemental 
superconductors, Hamlin {\it et al.}\cite{hamlin}
recently reported that Y (Z=39)  superconducts at T$_c$=17K 
under 89 GPa pressure and 19.5 K at 115 GPa, with the trend suggesting
higher T$_c$ at higher pressure.
This result illustrates that heavier elements should not be neglected;
note that La (Z=57) has T$_c$ up to 13 K under pressure.\cite{La1,La2}
The superconductivity of La has been interpreted in terms of the
rapidly increasing density of states of $4f$ bands near Fermi level 
with increasing pressure, causing 
phonon softening and resulting stronger coupling under 
pressure.\cite{warrenLa,lanthanum2}  Such a
scenario would not apply to Y, since there are no $f$ bands on
the horizon there.
No full calculations of the phonon spectrum and electron-phonon 
coupling have been carried out for either Y or La to date.

La and Y are two of the few elemental transition metals to have 
T$_c$ above\cite{jss1,table} 10 K, and the case of Y is unusually
compelling, since its value of T$_c$ is at least as high that of Li,
qualifying it as having
the highest T$_c$ of any elemental superconductor.  Moreover, the reduced
volume $v\equiv V/V_0$=0.42 corresponds to the value of T$_c \approx$ 20K
in Y [\onlinecite{hamlin}] (115 GPa) and also to the report of 
$T_c \approx$ 20K in (strained) Li  [\onlinecite{lithium1}] 
above 50 GPa.\cite{hanfland1,hanfland2}
For our study of Y reported here, it is first necessary to understand its
phase diagram.  Under pressure, it follows a
structure sequence\cite{vohra,grossans} through close-packed
phases that is typical of rare earth metals:
hcp$\to$Sm-type$\to$dhcp$\to$dfcc 
(dfcc is distorted fcc, with trigonal
symmetry).  The transitions occur around 12 GPa, 25 GPa, and 30-35 GPa.
Superconductivity was first found\cite{wittig} in Y by Wittig in the 11-17 GPa
range (1.2-2.8 K), in what is now known to be the Sm-type structure.
From 33 GPa to 90 GPa T$_c$ increases smoothly (in fact T$_c$ increases 
linearly with decrease in $v$ over the entire 35-90 GPa range\cite{hamlin}) 
suggesting that Y remains in the fcc phase, perhaps with the
distortion in the dfcc phase vanishing (the tendency is for the
$c/a$ ratio in these structures to approach ideal at high 
pressure\cite{vohra}).  
Calculations\cite{melsen} predict it adopts the bcc structure at
extremely high pressure($>$280 GPa), but this is far beyond our interest
here.

In this paper we report electronic structure and electron-phonon 
coupling calculations of Y 
for reduced volumes in the range 0.6$ \leq v \leq 1$ 
(pressures up to 42 GPa).
Our results indeed show strong electron-phonon coupling and phonon 
softening with increasing pressure.  A lattice 
instability (in the harmonic approximation used in linear response
calculations) is encountered at $v$=0.6 and persists to higher
pressures.  The instability arises from
the vanishing of the restoring force for transverse displacements for 
Q$\parallel <111>$ near the zone boundary, corresponding to sliding of
neighboring close-packed layers of atoms.  It is only the stacking
sequence of these layers that distinguishes the various structures in
the pressure sequence of structures (see above) that is observed in
rare earth metals.  Near-vanishing of the restoring force for sliding of
these layers is consistent with several stacking sequences being
quasi-degenerate, as the structural changes under pressure suggest.

This paper is organized as follows.  In Sec. II structural details are 
given, and the calculational methods are described.  Results for the 
electronic structure and its evolution with pressure are provided in
Sec. III.  The background for understanding the electron-phonon coupling 
calculations is provided in Sec. IV, and corresponding results are presented and
analyzed in Sec. V.  The implications are summarized in Sec. VI.

\section{Structure and calculation details}
Yttrium crystallizes in the hcp structure at ambient pressure with space 
group $P63/mmc$ (\#194) and lattice constants a=3.647~\AA~and 
c=5.731~\AA\cite{lattice}. 
 
Since the observed structures are
all close packed (or small variations from) and above 35 GPa Y is
essentially fcc, we
reduce the calculational task by using the fcc structure 
throughout our calculations. The space group is $Fm3m$ (\#225), with 
the equivalent ambient pressure lattice 
constant a=5.092~\AA.  We do note however that results 
for electron-phonon strength can be
sensitive to the crystal symmetry, both through the density of states
and through the nesting function that is described below.

We use the full potential local orbital (FPLO) code\cite{fplo} to study the
electronic structure, and apply the 
full-potential linear-muffin-tin-orbital (LMTART) code\cite{lmtart} 
to calculate the phonon frequencies and the electron-phonon coupling 
spectral function $\alpha^2 F$.
For FPLO, a k mesh of $36^3$ and the Perdew-Wang (PW92)\cite{pw92} 
exchange-correlation potential are used.
The basis set is 1s2s2p3s3p3d::(4s4p)/5s5p4d+. 
%The Fermi surfaces are obtained by using 
%the code XFSF. 
For LMTART, a k mesh of $48^3$ and GGA96 approximation\cite{gga96} for
exchange-correlation potential are used. For the electron-phonon coupling 
calculations we used a phonon Q 
mesh of 16$^3$, 
which has 145 Q points in the 
irreducible Brillouin zone.

\begin{figure}[tbp]
%\rotatebox{-90}
{\resizebox{8.2cm}{8.2cm}{\includegraphics{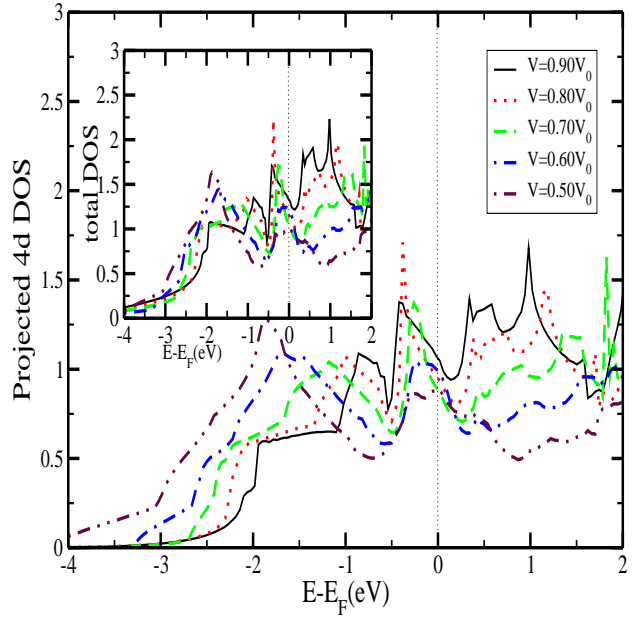}}}
\caption{(Color online)
Plot of the total DOS and projected $4d$ DOS per atom
of fcc Y with different
volumes. Both the total and the $4d$ density of states at Fermi level
decrease with reduction in volume.
}
\label{4ddos}
\end{figure}

\section{Electronic Structure under Pressure}
Many studies suggest that the general character of an elemental rare earth
metal is influenced strongly  by the occupation number
of the $d$ electrons, which changes under pressure. Our calculations show that
the $4d$ occupation number of trivalent Y increases
from 1.75 at ambient pressure, to a little above 2 at V=0.7V$_0$ and
then finally close to 3 at V=0.3V$_0$ (which is extreme pressure). 
Such an increase can be seen
from the projected density of states (PDOS) of $4d$ states (Fig. \ref{4ddos})
at different volumes.
From Fig. \ref{4ddos} broadening of the density of states with reduction
in volume can be seen, but is not a drastic effect.  The main
occupied $4d$ PDOS widens from 2 eV to 3 eV with reduction of the 
volume to $v$=0.5.  The value of of the density of states at the
Fermi level (taken as the zero of energy) $N(0)$ decreases 
irregularly with volume
reduction; the values are given in Table \ref{table}.

\begin{figure}[tbp]
%\rotatebox{-90}
{\resizebox{8.2cm}{6.0cm}{\includegraphics{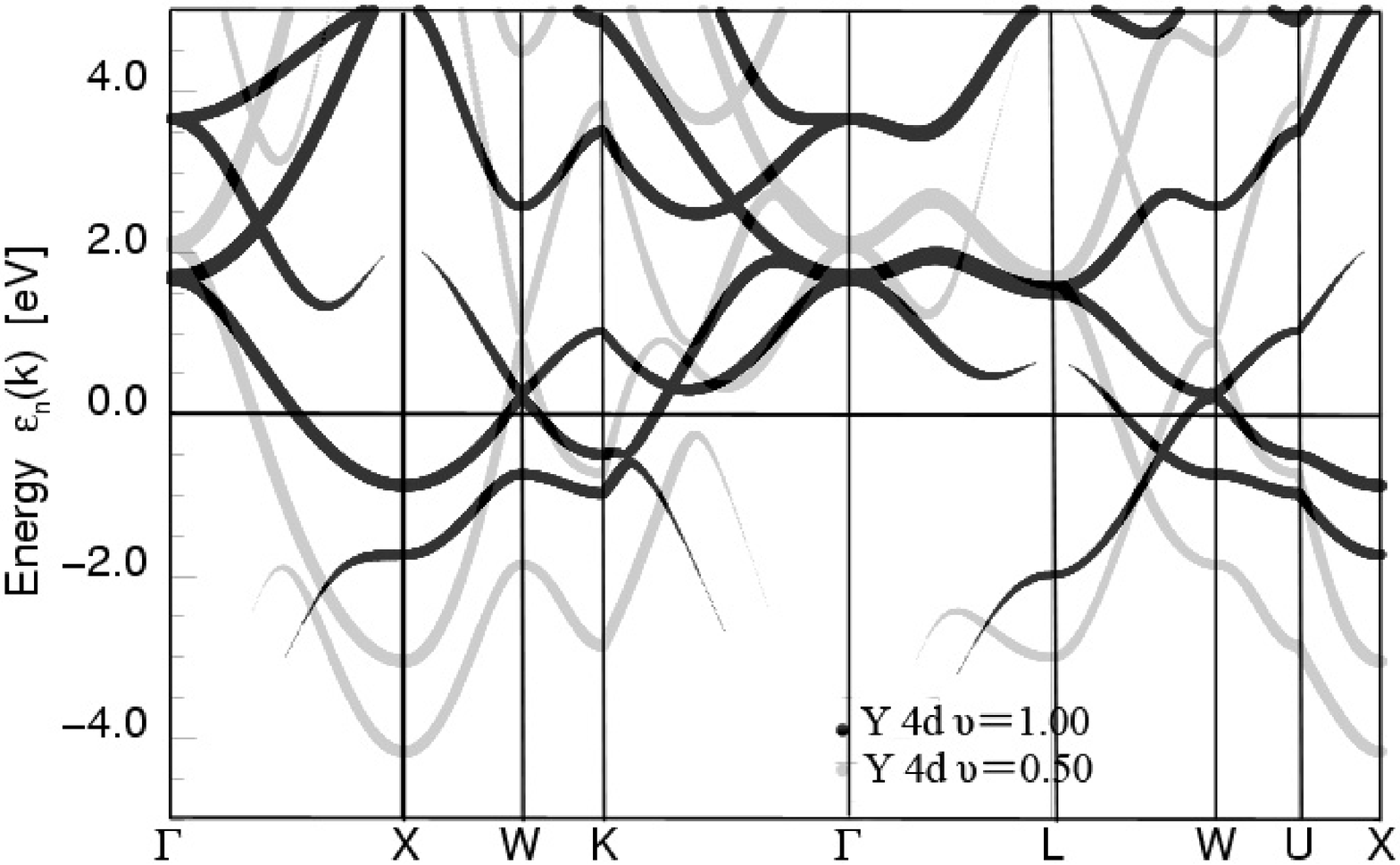}}}
\caption{(Color online)
Plot along high symmetry directions of the bands of Y at $V/V_o$=1.00
and at $V/V_o$=0.50.  The ``fattening'' of the bands is proportional
to the amount of Y $4d$ character.  Note that the $4d$ character
goes substantially in the occupied bands under pressure (the 
lighter shading), although
there is relatively little change in the Fermi surface band crossings.
}
\label{Fat}
\end{figure}

The pressure evolution of the band structure is indicated in Fig.
\ref{Fat}, where the $4d$ character at $v$=1.00 (black)
and $v$=0.50 (gray)
is emphasized.  First, the relative positions of the Fermi level
crossings change smoothly, indicating there is little change in
the Fermi surface topology.  This slow change is also seen in
Fermi surface plots, of which we show one (below).  Second, the 
overall band widths change moderately, as was noted above in the
discussion of the density of states.  The change in position of
$4d$ character is more substantial, however.  $4d$ bands at X lying
at -1 eV and -2 eV at ambient volume are lowered to -3 eV and -4 eV
at $v$=0.50.  Lowering of $4d$ character bands at K and W is
also substantial.  Thus Y is showing the same trends as seen in
alkali metals.  For Cs and related alkalies and alkaline earths 
under pressure, $6s$ character diminishes
as $5d$ character grows strongly with pressure.\cite{AKM}  In Li, $2s$
character at the Fermi surface evolves to strong $2p$ mixture\cite{Deepa}
at the volume where $T_c$ goes above 10 K.

\begin{figure}[tbp]
%\rotatebox{-90}
{\resizebox{7.2cm}{7.2cm}{\includegraphics{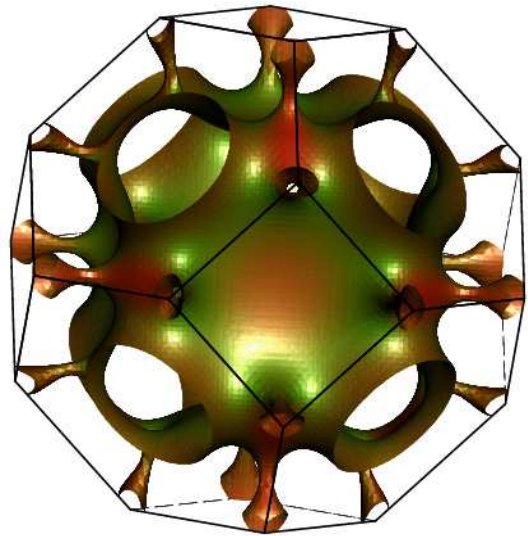}}}
\caption{(Color online)
Surface plot of the Fermi surface of fcc Y at a volume corresponding
to ambient pressure.  The surface is shaded according to the Fermi velocity.
The surface is isomorphic to that of Cu, {\it except} for the tubes
through the W point vertices that connect Fermi surfaces in neighboring 
Brillouin zones.  The evolution with pressure is described in the text. 
}
\label{FS}
\end{figure}

The Fermi surface of Y at ambient pressure (hcp) has been of interest
for some time, from the pioneering calculation of Loucks\cite{loucks}
to the recent measurements and calculations of Crowe {\it et al.}\cite{crowe}
However, the unusual Fermi surface in the hcp structure (having a
single strong nesting feature) is nothing like that
in the fcc phase we are addressing, which is unusual in its own way.  
At $v$=1.00 the fcc Fermi surface is a large `belly' connected by wide necks
along $<111>$ directions as in Cu, but in addition there are tubes
(`wormholes') connecting a belly to a neighboring zone's belly through 
each of the 24 W points.  The belly encloses holes rather than electrons
as in Cu; that is, the electrons are confined to a complex multiply-connected
web enclosing much of the surface of the Brillouin zone.

As the volume is reduced, the wormholes slowly grow in diameter until
in the range 0.5$< v <$0.6, they merge in certain places with the necks along
the $<111>$ directions, and the change in topology leaves closed 
surfaces around the K points as well as a different complex 
multiply-connected sheet.  The point we make is that, at all volumes, the
Fermi surface is very complex geometrically.  There is little hope
of identifying important ``nesting'' wavevectors short of an extensive
calculation.  Even for the simple Fermi surface of fcc Li, unexpected
nesting vectors were located\cite{Deepa} 
in three high symmetry planes of the zone.
The rest of the zone in Li still remains unexplored.

\section{Background: Electron-Phonon Coupling}
The electron-phonon spectral function $\alpha^{2}F(\omega)$ 
can be
expressed in terms of phonon properties [and N(0)] in the form\cite{alpha2F}
\begin{equation}\label{eq:alpha2F}
\alpha^{2}F(\omega)=\frac{1}{2\pi N(0)} 
   \sum\limits_{\mathbf{Q}\nu}\frac{\gamma_{\mathbf{Q\nu}}}
{\omega_{\mathbf{Q\nu}}}\delta(\omega-\omega_{\mathbf{Q\nu}})
\end{equation}
where 
\begin{eqnarray}
F(\omega)=\sum\limits_{\mathbf{Q}\nu}\delta(\omega-\omega_{\mathbf{Q}\nu}) 
\end{eqnarray}
is the density of phonon states, and
$N(0)$ is the 
single spin Fermi surface density of states. 
The phonon linewidth $\gamma_{\mathbf{Q}\nu}$ is given by
\begin{equation}\label{eq:gammaQ}
\gamma_{\mathbf{Q\nu}}=2\pi\omega_{\mathbf{Q\nu}}
          \sum\limits_{\mathbf{k}}
    |M_{\mathbf{k}+\mathbf{Q},\mathbf{k}}^{[\nu]}|^{2}
	  \delta(\varepsilon_{\mathbf{k}})
\delta(\varepsilon_{\mathbf{k}+\mathbf{Q}})
\end{equation}
where $M_{\mathbf{k}+\mathbf{Q}}^{[\nu]}$ 
is the electron-phonon matrix element; $\nu$ is the branch index.  
Sums over $\mathbf{Q}$ or 
$\mathbf{k}$ are conventionally normalized (divided by the number of
unit cells in the normalization volume).  

The quantities thus defined enable one to identify the contribution 
to $\lambda$ from each mode, the ``mode $\lambda$'', as
\begin{equation}\label{eq:lambdaQ}
\lambda_{\mathbf{Q}\nu}=\frac{2}{\omega_{_{\mathbf{Q}\nu}}
    N(0)}\sum\limits_{\mathbf{k}}
  |M_{\mathbf{k},\mathbf{k}+\mathbf{Q}}^{[\nu]}|^{2}
     \delta(\varepsilon_{\mathbf{k}})
\delta(\varepsilon_{\mathbf{k}+\mathbf{Q}})
\end{equation}
With this definition $\lambda$ is the 
average over the zone, and sum (not average) over the 
$N_{\nu}=3$ branches of all of the 
$\lambda_{\mathbf{Q}\nu}$ values.
The electron-phonon coupling strength $\lambda$ then is given by
\begin{eqnarray}\label{eq:lambda}
\lambda = \frac{4}{\pi N(0)}\sum\limits_{\mathbf{Q}\nu}
      \frac{\gamma_{\mathbf{Q}\nu}}{{\omega_{\mathbf{Q}\nu}}^2}
  \equiv \sum\limits_{\mathbf{Q}\nu} \lambda_{\mathbf{Q}\nu}.
\end{eqnarray}

The critical temperature
T$_c$ can be obtained by using the Allen-Dynes modification of the
McMillan formula,\cite{Allen, McMillan} which depends on the logarithmic,
first, and second frequency moments $\omega_{log}$, 
$\omega_1 \equiv <\omega>$, and
$\omega_2 \equiv <\omega^2>^{1/2}$, as well as
$\lambda$ and the Coulomb pseudopotential $\mu^*$.  
These averages are weighted according 
to the normalized coupling `shape function' 
2$\alpha^2 F(\omega)/(\lambda \omega)$.  They are often, and will be
for Y especially under pressure, much different from simple averages
over the spectrum F($\omega$).

Note that $\lambda_{\mathbf{Q}\nu}$, or $\gamma_{\mathbf{Q}\nu}$, 
incorporates a phase space factor,
the `nesting function'\cite{Deepa} $\xi({\mathbf Q})$ 
describing the phase space that is available for electron-hole
scattering across the Fermi surface(E$_F$=0),
\begin{equation}\label{xi}
\xi(\mathbf{Q})=\frac{1}{N}\sum\limits_{\mathbf{k}}
  \delta(\varepsilon_{\mathbf{k}})
  \delta(\varepsilon_{\mathbf{k}+\mathbf{Q}})\varpropto\oint_{\cal L} 
 \frac{d{\cal L}_{\mathbf{k}}}{|{\mathbf{v}}_{\mathbf{k}}\times
    {\vec v}_{\mathbf{k+Q}}|}.
\end{equation}
Here ${\cal L}$ is the line of intersection of an undisplaced
Fermi surface and one displaced by ${\mathbf{Q}}$, and ${\vec v}_k$
is the electron velocity at $\vec k$.
These equations presume the adiabatic limit, in which the phonon
frequencies are small compared to any electronic energy scale.  This
limit applies to elemental Y.

\section{Results and Analysis}

\begin{figure}[tbp]
%\rotatebox{-90}
{\resizebox{8.0cm}{8.0cm}{\includegraphics{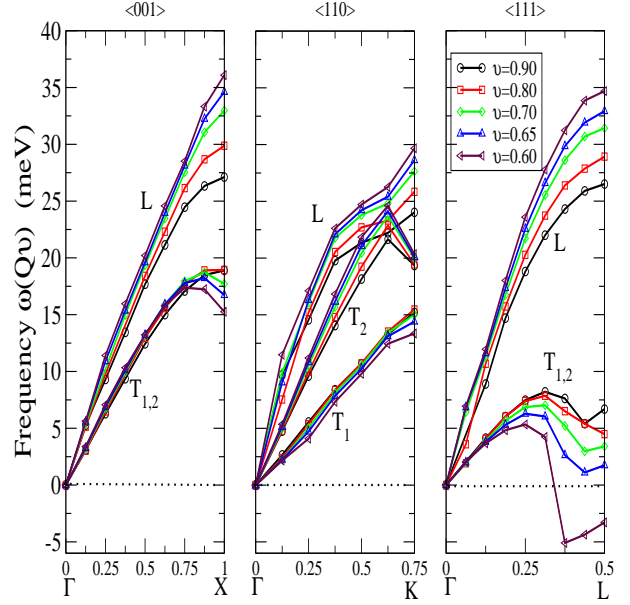}}}
\caption{(Color online)
Plot of the calculated phonon spectrum along high symmetry directions
($\Gamma$-X,
$\Gamma$-K, $\Gamma$-L) of fcc Y with different volumes. The longitudinal mode phonons increases with the distance
from $\Gamma$ points along all the three directions. 
Along $\Gamma$-X direction (left panel), the doubly degenerate 
traverse mode slightly softens near X point, 
while along $\Gamma$-K direction (left panel, only the T$_2$ mode 
sightly softens near K point.
Along $\Gamma$-L direction (right panel), the already soft doubly degenerate 
transverse mode soften further near the L point 
with decreasing volume. At $V=0.6V_0$, the frequency at L becomes 
negative, indicating lattice instability.   
   }
\label{phonons}
\end{figure}

\subsection{Behavior of Phonons}
The calculated phonon branches are shown along the high symmetry 
lines, from $v$=0.90
down to $v$=0.60, in Fig. \ref{phonons}.
The longitudinal modes behave normally, increasing monotonically
in frequency by $\sim30$\% in this range.  The transverse modes
along $<100>$ and $<110>$ show little change; the doubly degenerate
transverse mode at X {\it softens} by 20\%, reflecting some unusual
coupling.  Along $<110>$, T$_1$ and T$_2$ denote modes polarized 
in the $x-y$ plane, and along the $z$ axis, respectively.

The interesting behavior occurs for the (doubly degenerate) transverse
branch along $<111>$.  It is quite soft already at $v$=0.9 (7 meV, only 25\%
of the longitudinal branch), softer than the corresponding mode in hcp Y at
ambient pressure.\cite{sinha}  With decreasing volume it {\it softens} 
monotonically, and becomes unstable between $v$=0.65 and $v$=0.60.
It should not be surprising that the transverse mode at the L point is
soft in a rare earth metal.  The sequence of structural transitions
noted in the introduction (typically 
hcp$\to$Sm-type$\to$dhcp$\to$dfcc$\to$fcc for trivalent elements) 
involves only different stacking of
hexagonal layers of atoms along the cubic (111) direction.  So 
although these various periodic stackings may have similar energies, the soft
(becoming unstable) transverse mode at L indicates also that the barrier
against sliding of these planes of atoms is very small.  At $v$=0.60
(see Fig. \ref{phonons}) the largest instability is not at L itself
but one-quarter of the distance back toward $\Gamma$.  At $v$=0.65
there are surely already anharmonic corrections to the lattice
dynamics and coupling from the short wavelength transverse branches.

\begin{figure}[tbp]
%\rotatebox{-90}
{\resizebox{8.2cm}{8.2cm}{\includegraphics{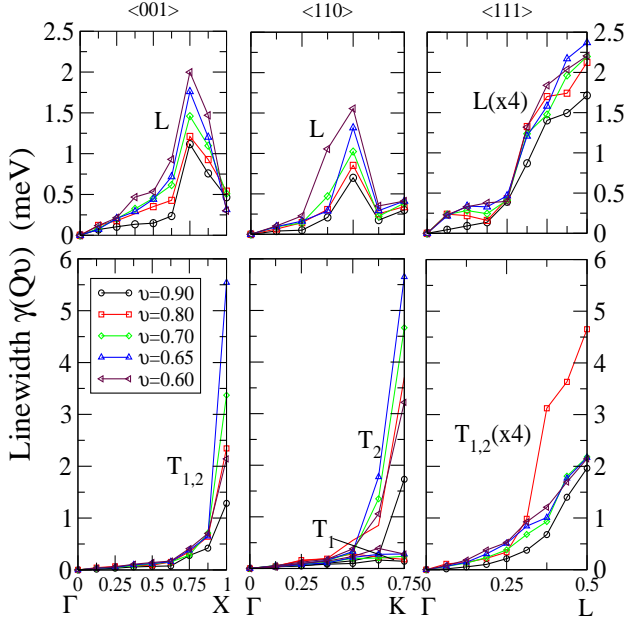}}}
\caption{(Color online)
Plot of the calculated linewidths of fcc Y  for varying 
volumes. The linewidths of the transverse modes at the X point increases 
from 1.3
to 5.5 as volume decreases from V=0.9V$_0$ to V=0.6V$_0$. 
The linewidths of the T$_2$ along $<110>$ modes show 
the same increase.  The linewidths along the $<111>$ direction have
been multiplied by four for clarity.
%The linewidths of the longitudinal modes at
%Q point (0,0,3/4) and (1/2, 1/2, 0) also show similar behavior. 
   }
\label{linewidth}
\end{figure}

\subsection{Linewidths}
The linewidths $\gamma_{Q\nu}$, one indicator of 
the mode-specific contribution to T$_c$,
are shown in Fig. \ref{linewidth}. To understand renormalization,
one should recognize that in lattice dynamical
theory it is $\omega^2$, and not $\omega$ itself, that arises
naturally.  At $v$=0.90, $\omega^2$ for the T modes is only 1/16 of
the value for the longitudinal mode at the L point.  A given amount 
of coupling will affect the soft modes much more strongly than it
does the hard modes.

For the $<110>$ direction, the strong peak in $\gamma_{Q\nu}$ for the T$_2$
$\hat z$ polarized) mode at the zone boundary point K 
(5.7 meV) is reflected in the 
dip in this mode at K that can be seen in Fig. \ref{phonons}.   
At $v$=0.60 the linewidth is 1/3 of the frequency.  The coupling to the
T$_1$ mode along this line is negligible.  Note that it is the T$_1$
mode that is strongly coupled in Li and is the first phonon to
become unstable.  A peak in the linewidth of the L modes correlates
with a depression of the frequency along this line.
Along $<001>$ the T modes again acquire large linewidths near the
zone boundary under pressure.  This electron-phonon coupling is
correlated with the dip in the T frequency in the same region.

The coupling along the $<111>$ direction is not so large, either
for T or for L branches (note, they have been multiplied by a factor
of four in Fig. \ref{linewidth}.  The coupling is strongest at the
zone boundary, and coupled with the softness already at $v$=0.90,
the additional coupling causes an instability when the volume is
reduced to $v$=0.60
(P = 42 GPa).  This seems to represent an example where a rather
modest amount of coupling has a potentially catastrophic result:
instability of the crystal.  Evidently Y is stabilized in the fcc
structure by anharmonic effects, coupled with the fact that being
already close-packed there may be no simple structural phase that 
is lower in energy.

\begin{figure}[tbp]
%\rotatebox{-90}
{\resizebox{8.2cm}{8.2cm}{\includegraphics{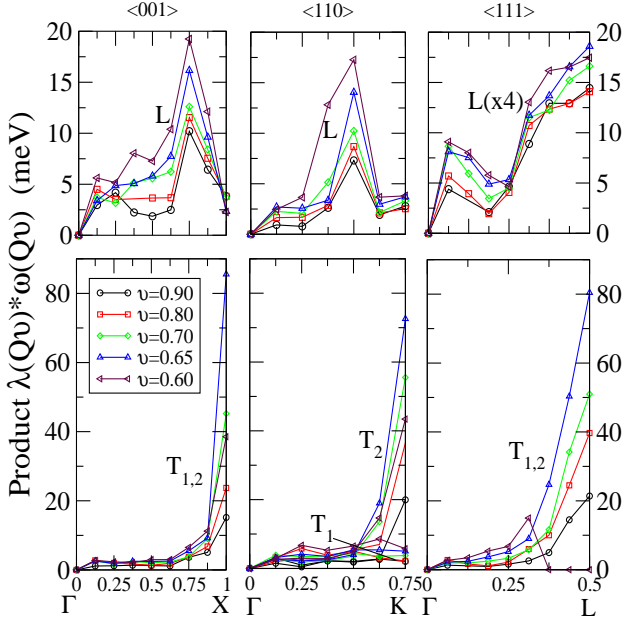}}}
\caption{(Color online)
Plot of the product $\lambda_{Q\nu} \omega_{Q\nu}$ of fcc Y for different
volumes, along the high symmetry directions.  Note that the longitudinal
(L) values along $<111>$ have been multiplied by four for clarity.
In addition, values corresponding to unstable modes near L have been
set to zero.  Differences in this product reflect differences in
matrix elements; see text.
   }
\label{product}
\end{figure}

\subsection{Coupling Strength}
It is intuitively clear that strong coupling to extremely low frequency
modes is not as productive in producing high T$_c$ as coupling to
higher frequency modes.  This relationship was clarified by
Bergmann and Rainer,\cite{rainer} who calculated the functional
derivative $\delta T_c/\delta \alpha^2 F(\omega)$.  They found that
coupling at frequencies less than $\bar \omega$
$= 2\pi$T$_c$ has little impact
on T$_c$ (although coupling is never harmful).  Since we are thinking
in terms of Y's maximum observed T$_c \approx$ 20 K, this means
that coupling below $\bar \omega$ = 10 meV becomes ineffective.

The product $\lambda_{Q\nu} \omega_{Q\nu} \propto 
\gamma_{Q\nu}/\omega_{Q\nu}$ gives a somewhat different indication
of the relative coupling strength\cite{carbotte} 
than does either $\lambda_{Q\nu}$
or $\gamma_{Q\nu}$.  It is also, up to an overall constant, just
the nesting function defined earlier, with electron-phonon matrix
elements included within the sum.  Since the nesting function is a
reflection of the phase space for scattering, it is independent of the
polarization of the mode, hence differences between the three branches
are due solely to the matrix elements.

This product $\lambda_{Q\nu} \omega_{Q\nu}$ is shown in Fig. \ref{product}.
The weight in the transverse modes is concentrated near the zone
boundary, with the region being broader around L than at X or K and
growing in width with pressure.
The $T_1$ branch along $<110>$, which is polarized along [$1\bar{1}0$], shows
essentially no coupling.  The weight in this product for the longitudinal
modes is peaked {\it inside} the zone boundary along the $<001>$ and
$<110>$ directions with a mean value of 7-8 meV, and is weaker 
along the $<111>$ direction.  

\subsection{$\alpha^2 F(\omega)$}
\begin{figure}[tbp]
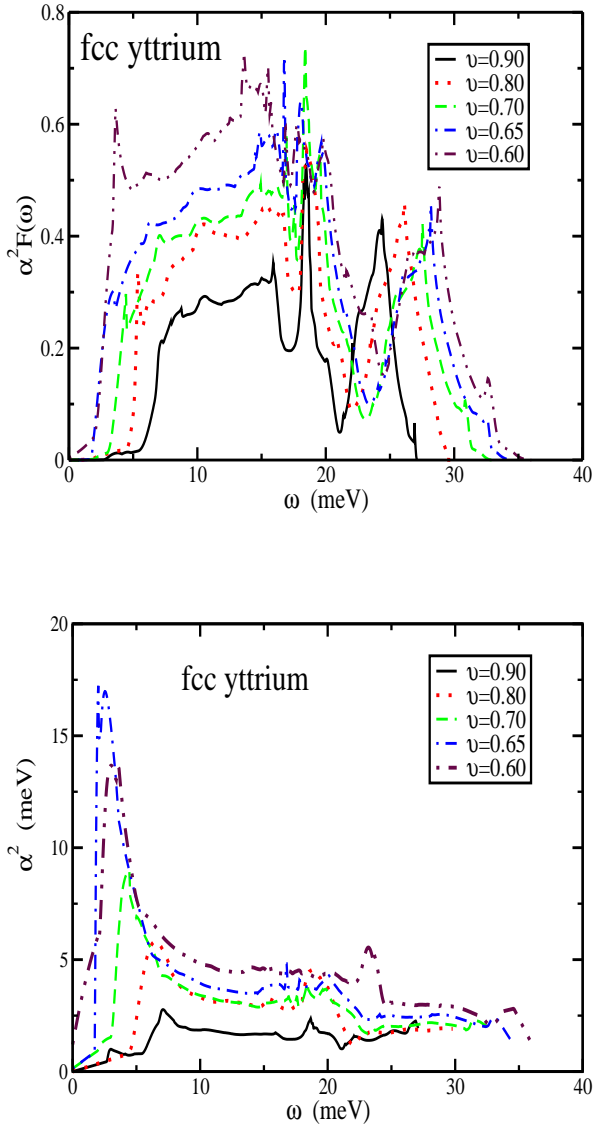

%\rotatebox{-90}
{\resizebox{7.8cm}{6.8cm}{\includegraphics{fig7.eps}}}
\vskip 13mm
{\resizebox{7.8cm}{6.8cm}{\includegraphics{fig8.eps}}}
\caption{(Color online) Top panel:
Plot of $\alpha^2F(\omega)$ versus $\omega$ As volume decreases,  $\alpha^2F(\omega)$ increases and gradually 
transfers to low frequency.
Bottom panel: the frequency-resolved coupling strength $\alpha^2(\omega)$
for each of the volumes studied.  The evolution with increased pressure
is dominated by strongly enhanced coupling at very low frequency
(2-5 meV).
   }
\label{a2F}
\end{figure}

The results for $\alpha^2F$ are displayed in Fig. \ref{a2F}.
The longitudinal peak in the 20-35 range hardens normally with
little change in coupling strength.  The 7-20 meV range of transverse
modes at $v$=0.90 increases in width to 2-24 meV at $v$=0.60,
and the strength increases monotonically and strongly.  The
strong peak in $\alpha^2(\omega)= \alpha^2F(\omega)/F(\omega)$, 
shown in the bottom panel of Fig. \ref{a2F},
reflects the very soft modes that have been driven into the 2-5 meV
range, and the fact that they are very relatively strongly coupled.
The substantial increase in coupling, by a factor of $\sim$2.5,
in the range 7-25 meV is important for T$_c$, as noted in the
next subsection.

\begin{table}
\caption{For each volume $v$ studied, the columns give the
experimental pressure\cite{melsen}(GPa), the Fermi level density
of states N(0) (states/Ry spin),
and calculated values of the mean frequency $\omega_1 = <\omega>$ (meV),
the logarithmic moment $\omega_{log}$ and second moment
$\omega_2 = <\omega^2>^{1/2}$ (all in meV),
the value of $\lambda$, the product $\lambda\omega_2^2$ (meV$^2$),
and T$_c$ (K).  For T$_c$ the value of the
Coulomb pseudopotential was taken as $\mu^*$=0.15.}
\begin{tabular}{|r|r|r|r|r|r|r|r|r|}
\hline
v & P &N(0)&$\omega_{log}$&$\omega_1$& $\omega_2$ & $\lambda$ &$\lambda \omega_2^2$& T$_c$\\
 \hline
 0.90     &~6  &  9.7 &12.5& 13.6    & 14.7   & 0.75  &162    & ~~4.0 \\
 0.80     &14  & 11.3 &11.6& 13.2    & 14.5   & 1.30  &273    & ~11.9 \\
 0.70     &26  &  9.1 &10.1& 12.0    & 13.8   & 1.53  &291    & ~13.0 \\
 0.65     &32  &  8.4 & 7.6& 10.2    & 12.6   & 2.15  &341    & ~14.4 \\
 0.60     &42  &  7.9 & 6.9& ~9.5    & 12.1   & 2.80  &410    & (16.9) \\
 \hline
 \end{tabular}
 \label{table}
 \end{table}

\subsection{Estimates of T$_c$}
This background helps in understanding the trends displayed in
Table \ref{table}, where T$_c$ from the Allen-Dynes equation\cite{Allen}
(choosing the standard value of $\mu^*$=0.15)
and the contributing materials
constants are displayed.  The calculated values of $\lambda$ 
increases strongly, by a factor of
3.7 in the volume range we have studied.  Between $v$=0.65 and
$v$=0.60 (the unstable modes are removed from consideration)
$\lambda$ increases 30\% but T$_c$ increases by only 2.5 degrees.
The cause becomes clear in looking at the frequency moments.  
These moments are weighted by $\alpha^2F(\omega)/\omega$. 
$\alpha^2(\omega)$ itself become strongly peaked at low frequency under pressure,
and it is further weighted by $\omega^{-1}$. The frequency moments
set the scale for T$_c$ ($\omega_{log}$ in particular) and 
they {\it decrease} strongly with decreasing volume.  In particular,
$\omega_{log}$ decreases by 45\% over the volume range we have
studied, reflecting its strong sensitivity to soft modes.  

The increase in T$_c$ probably owes more to the increase
in coupling in the 10-25 meV range (see $\alpha^2(\omega)$ plot in
Fig. \ref{a2F}; a factor of roughly 2.5) than to the more
spectacular looking peak at very low frequency.  Put another way,
the very low frequency peak in $\alpha^2F$ looks impressive and
certainly contributes strongly to $\lambda$, but is also very effective
in lowering the temperature scale ($\omega_{log}$).
For $\alpha^2(\omega)$ shapes such as we find for Y, the quantities
$\lambda$ and $\omega_{log}$ which go into the Allen-Dynes equation
for T$_c$ do not provide a very physical picture of the change in 
T$_c$.  For this reason we provide also in Table \ref{table} the
product $\lambda \omega_2^2 = N(0)<I^2>/M$ ($<I^2>$ is the conventional
Fermi surface average of square of the electron-ion matrix element
and $M$ is the atomic mass).  For the volumes 0.60$\leq v \leq$0.80
in the table, the ratio of $\lambda \omega_2^2$/T$_c$ is nearly
constant at 23$\pm$1.5 (in the units of the table), illustrating 
the strong cancellation of the increase of $\lambda$ with the
decrease in frequency moments in producing the resulting T$_c$.

\section{Summary}
In this paper we have presented the evolution of elemental Y over
a range of volumes ranging from low pressure to 40+ GPa pressure
($V/V_o$ = 0.60).  
Lattice instabilities that emerge near this pressure (and persist
to higher pressures) make
calculations for smaller volumes/higher pressures unrealistic.  For simplicity
in observing trends the structure has been kept cubic (fcc); however,
the observed phases are also close-packed so it was expected
that this restriction may
still allow us to obtain the fundamental behavior underlying the
unexpectedly high T$_c$ in Y.  On the other hand, the Fermi surface
geometry varies strongly with crystal structure, and the nesting
function $\xi(Q)$ and perhaps also the matrix elements may have
some sensitivity to the type of long-range periodicity.  

In addition to the band structure, Fermi surface, and electronic 
density of states, we have also presented the phonon dispersion
curves and linewidths along the high symmetry directions, and also
have presented $\alpha^2 F(\omega)$ and the resulting value of T$_c$.
The results show indeed that Y under pressure becomes a
strongly coupled electron-phonon system, readily accounting for value
of T$_c$ in the range that is observed.  

In spite of having used a relatively dense mesh of Q points for the
phonons, it seems clear that this Brillouin zone integral is still not
well converged.  Evaluation of $\xi(Q)$ on a very fine Q mesh in 
three planes for fcc Li, which has a very simple Fermi surface, has
shown\cite{Deepa} that this nesting factor contains (thickened)
surfaces of fine structure with high intensity.  The convergence 
of this zone integral (and for example the resulting $\alpha^2F$ function)
has rarely been tested carefully in full linear response evaluations
of phonons; such a test could be very computationally intensive.
Nevertheless, the general finding of strong coupling is clear.

Very recently it has been found that isovalent 
Sc is superconducting at 8.1 K under 74 GPa pressure.\cite{hamlin2}
Note that if the lattice were harmonic and the only difference 
between Sc and Y were the masses (which differ by a factor of two),
T$_c$ = 20 K for Y would translate to T$_c$ = 28 K for Sc. (For an
element with a harmonic lattice, $\lambda$ is independent of mass.)
The corresponding argument for (again isovalent) La gives T$_c$ = 16 K.
La has T$_c$ = 13 K at 15 GPa, and has not been studied 
beyond\cite{jss1} 45 GPa.

Another comparison may be instructive.  Dynes and Rowell obtained and
analyzed tunneling data\cite{rowell} on 
Pb-Bi alloys where $\lambda$ is well into
the strong coupling region, becoming larger than two as is the
case for Y under pressure in Table
\ref{table}.  The Pb$_{0.65}$Bi$_{0.35}$ alloy has $\lambda$=2.13,
$<\omega^2>$=22.6 meV$^2$.
We can compare directly with the $v$=0.65 case in Table \ref{table},
which has $\lambda$=2.15, $<\omega^2>$=159 meV$^2$.  The product
$M<\omega^2>$ for Y is three times as large as for the heavy alloy.
Since the $\lambda$'s are equal, the value of $N(0)<I^2>$ 
(equal to $\lambda M <\omega^2>$) is 
also three times as large as in the alloy.  The values of T$_c$ are
14.4 K (Y) and 9 K (alloy) [somewhat different values of $\mu^*$
were used.]  The values of $\omega_{log}$ differ by less than a factor
of two, due to the low-frequency coupling in $\alpha^2(\omega)$ 
in Y that brings
that frequency down, and that is why the values of T$_c$ also differ
by less than a factor of two.

While this study is in some sense a success, in that it has become
clear that strong electron-phonon coupling can account for the 
remarkable superconductivity of Y under pressure, there remains a
serious shortcoming, one that is beyond the simple lack of 
numerical convergence that would pin down precisely $\lambda$,
T$_c$, etc.  What is lacking is even a rudimentary physical picture
for what distinguishes Y and Li (T$_c$ around 20 K under pressure) from
other elemental metals which show low, or vanishingly small, values
of T$_c$.  

The rigid muffin-tin approximation (RMTA) of Gaspari and 
Gyorffy,\cite{GG} which approximates
the phonon-induced change in potential and uses an isotropic idealization
for the band structure to derive a simple result, seemed fairly 
realistic for the electronic contribution (the Hopfield $\eta$)
for transition metal elements and intermetallics.\cite{Z50,wep}  On top of
these idealizations, there is an additional uncertainty in 
$<\omega^2>$ that must be guessed to obtain $\lambda$ and T$_c$.
One would not `guess' the values of the frequency moments that we
have obtained for Y under pressure.

In addition, the RMTA expression does not distinguish between 
the very different matrix
elements for the various branches, giving only a polarization and
Fermi surface average.  Nevertheless, it gave a very useful 
understanding of trends\cite{Z50} in electron-phonon coupling in elemental
transition metals and in some intermetallic compounds. 
While the linear response evaluation
of the phonon spectrum and the resulting coupling seems to work well,
this more detailed approach has not
yet provided -- even for elemental superconductors -- the physical
picture and simple trends that would enable us to claim that we
have a clear understanding of strong coupling superconductivity.

\section{Acknowledgments}
We have benefited from substantial exchange of information 
with J. S. Schilling,
and help with computer codes from D. Kasinathan.
Z.P.Y. and W.E.P. were supported by National Science Foundation Grant No.
DMR-0421810. S.Y.S. acknowledges support from National Science 
Foundation Grants
DMR-0608283 and DMR-0604531.
W.E.P. is grateful for support from the Alexander von Humboldt Foundation,
and the hospitality of IFW Dresden, during the 
preparation of this manuscript.

\end{document}